\documentclass[10pt]{amsart}
\usepackage{amsmath, color, multirow}
\usepackage{amssymb}
\usepackage{amsfonts}
\usepackage{amsthm}
\usepackage{mathrsfs}
\usepackage[all]{xy}

\newcommand{\be}{\begin{equation}}
\newcommand{\ee}{\end{equation}}
\newtheorem{theorem}{Theorem}[section]

\theoremstyle{definition}

\theoremstyle{remark}
\newtheorem{remark}[theorem]{Remark}

\numberwithin{equation}{section}

\begin{document}

\title{Comment on ``One-way deficit of two qubit X states''}
\author{Naihuan Jing, Xia Zhang$^*$, Yao-Kun Wang}
\address{XZ: School of Mathematics, South China University of Technology,
Guangzhou, Guangdong 510640, China}
\email{1134319889@qq.com}
\address{NJ: Department of Mathematics, North Carolina State University, Raleigh, NC 27695, USA}
\email{jing@ncsu.edu}
\address{YW: College of Mathematics, Tonghua Normal University, Tonghua, Jilin 134001, China}
\email{wangyaokun521@126.com}

\thanks{{\scriptsize
\hskip -0.4 true cm MSC (2010): Primary: 81P40; Secondary: 81Qxx.
$*$Corresponding author, }} 

\begin{abstract}
We improve the recent method of Wang et. al to calculate exactly
the one-way information deficit of any X-state. Analytical formulas of the one-way information deficit are given for several nontrivial regions of the parameters.
\end{abstract}

\maketitle

\section{Introduction}

Quantum discord is one of the quantum correlations that has been studied intensively similar to
quantum entanglement in
quantum information process. It measures the difference between the mutual information and maximal classical
mutual information \cite{HHHO}. Besides the quantum discord, other quantum correlations
have also been introduced in various applications \cite{OHHH, OHHH2}, for example, one-way information deficits.
In \cite{S, C}, Streltsov et al. defined the quantity by the relative entropy, which has shown its
close relationship with quantum entanglement. The one-way information deficit by the Von Neumann measurement
for a quantum density matrix $\rho^{ab}$ with respect to the first component is given by \cite{S2}
\begin{equation}\label{e:def}
\Delta^\rightarrow(\rho^{ab})=\min_{\{P_i^a\}}S(\sum_iP_i^a\rho^{ab}P_i^a)-S(
\rho^{ab}),
\end{equation}
where the minimum is taken over Von Neuannn measurements and $S$ is the entropy. The Von Neuannn measurements
are parametrized by the set of all complete rank one orthogonal projectors
$P_i^a=|i^a\rangle\langle i^a|$ such that $\sum_i P_i^a=I$. In general it is a complex problem to compute the one-way information deficit exactly,
similar to the situation of the quantum discord (cf. \cite{L}).

Recently in \cite{Wang2}, Wang et al. have proposed a method to evaluate the one-way information deficit for an $X$ state,
which generalizes the earlier work \cite{Wang}. The idea of this new method is to reduce the
calculation to an optimization question with fewer variables. However, we find that Ref. \cite{Wang2} provides
the answer for degenerate cases and in general only gives an estimate.
In this note, we generalize and improve their method and obtain
an exact answer for the one-way information deficit of any $X$-state. Furthermore, several
analytical formulas of the one-way information deficit are given
for some nontrivial regions of the parameters, in particular, these include
cases with nonzero $z$-components in the Bloch decomposition of the density matrix.

\section{one-way information deficit for X states}
First of all, one observes that the one-way information deficit is invariant under local unitary transformations.
For a general two qubit X-state one considers the quantum state in the following form:
\begin{equation}\label{Eq2.1}
\rho^{ab}=\frac{1}{4}(I\otimes I+\sum_{i=1}^{3}r_i\sigma_i\otimes I+I\otimes \sum_{i=1}^3s_i\sigma_i+\sum_{i=1}^3c_i\sigma_i\otimes \sigma_i),
\end{equation}
where $\textbf{r}=\{r_1,r_2,r_3\}$, $\textbf{s}=\{s_1,s_2,s_3\}$ are the Bloch vectors,
$\sum_ir_i^2\leq 1, \sum_is_i^2\leq 1, \sum_ic_i^2\leq 1$ and $\{\sigma_i\}_{i=1}^3$ are the standard
Pauli spin matrices.
Here the Bloch vectors are assumed in the $z$-direction, that is,
$\textbf{r}=\{0,0,r_3\}$, $\textbf{s}=\{0,0,s_3\}$. Then the state can be written as
\begin{equation}
\rho^{ab}=\frac{1}{4}(I\otimes I+r_3\sigma_3\otimes I+s_3I\otimes\sigma_3+\sum_{i=1}^3c_i\sigma_i\otimes \sigma_i),
\end{equation}
where $r_3, s_3\in [-1, 1]$ and $\sum_ic_i^2\leq 1$.
In terms of the computational basis $|00\rangle$, $|01\rangle$, $|10\rangle$, $|11\rangle$, the matrix form of $\rho^{ab}$ is
\begin{equation}\label{Eq2.3}
\rho=\frac{1}{4}\begin{pmatrix}
     1+r_3+s_3-c_3 & 0 & 0 & c_1-c_2 \\
     0 & 1+r_3-s_3-c_3 & c_1+c_2 & 0 \\
     0 & c_1+c_2 & 1-r_3+s_3-c_3 & 0 \\
     c_1-c_2 & 0 & 0 & 1-r_3-s_3+c_3 \\
   \end{pmatrix}.\end{equation}
The eigenvalues are:
\begin{equation*}
\eta_{1,2}=\frac{1}{4}(1-c_3\pm\sqrt{(r_3-s_3)^2+(c_1+c_2)^2}),
\end{equation*}
\begin{equation*}
\eta_{3,4}=\frac{1}{4}(1+c_3\pm\sqrt{(r_3-s_3)^2+(c_1-c_2)^2}).
\end{equation*}
Then the entropy of $\rho$ is given by
\begin{align}\nonumber
S(\rho)=2&-\frac{1}{4}(1-c_3+\sqrt{(r_3-s_3)^2+(c_1+c_2)^2})\log_2(1-c_3+\sqrt{(r_3-s_3)^2+(c_1+c_2)^2})\\ \nonumber
&-\frac{1}{4}(1-c_3-\sqrt{(r_3-s_3)^2+(c_1+c_2)^2})\log_2(1-c_3-\sqrt{(r_3-s_3)^2+(c_1+c_2)^2})\\ \label{Eq2.4}
&-\frac{1}{4}(1+c_3+\sqrt{(r_3-s_3)^2+(c_1-c_2)^2})\log_2(1+c_3+\sqrt{(r_3-s_3)^2+(c_1-c_2)^2})\\ \nonumber
&-\frac{1}{4}(1+c_3-\sqrt{(r_3-s_3)^2+(c_1-c_2)^2})\log_2(1+c_3-\sqrt{(r_3-s_3)^2+(c_1-c_2)^2}).
\end{align}

Note that any Von Neumann measurement can be realized by
\begin{equation}
\{B_k=V\Pi_kV^\dag: k=0,1\}
\end{equation}
where $\Pi_k=|k\rangle\langle k|: k=0,1$ and $V\in \mathrm{SU}(2)$. Then the minimum in Eq. (\ref{e:def})
is taken over the group $\mathrm{SU}(2)$.  We will compute the
one-way information deficit with respect to the second particle and the superscript
of $\Pi_k^b$ is removed for simplicity.
Each unitary operator $V\in \mathrm{SU}(2)$
can be written as $$V=tI+i\sum_{i=1}^3y_i\sigma_i$$ where $t, y_i\in \mathbb{R}$
and $t^2+\sum_{i=1}^3y_i^2=1.$

After the measurement $\{B_k\}$, the state $\rho$ is changed to the ensemble $\{\rho_k, p_k\}$:
\begin{equation}
\rho_k=\frac{1}{p_k}(I\otimes B_k)\rho(I\otimes B_k),
\end{equation}
where $p_k=tr(I\otimes B_k)\rho(I\otimes B_k)$. It follows from
\cite{Wang2} that
\begin{align*}
p_0\rho_0&=\frac{1}{4}(I+s_3z_3I+r_3\sigma_3+\sum_{i=1}^3(c_iz_i)\sigma_i)\otimes(V\Pi_0V^\dag),\\
p_1\rho_1&=\frac{1}{4}(I-s_3z_3I+r_3\sigma_3-\sum_{i=1}^3(c_iz_i)\sigma_i)\otimes(V\Pi_1V^\dag),
\end{align*}
where $z_1=2(y_1y_3-ty_2)$, $z_2=2(ty_1+y_2y_3)$, $z_3=t^2+y_3^2-y_1^2-y_2^2$.
Here it is easy to check directly that $\sum_{i=1}^3 z_i^2=1$. So the minimum in the one-way
information deficit is taken over the unit sphere. 

The eigenvalues of $p_0\rho_0+p_1\rho_1$ are
\begin{align*}
\lambda_{1,2}&=\frac{1}{4}\left(1+s_3z_3\pm\sqrt{r^2_3+2r_3z_3+\sum_i(c_iz_i)^2}\right),\\
 \lambda_{3,4}&=\frac{1}{4}\left(1-s_3z_3\pm\sqrt{r^2_3-2r_3z_3+\sum_i(c_iz_i)^2}\right).
\end{align*}

Let $\phi=z_3$, $\theta=c_1^2z_1^2+c_2^2z_2^2+c_3^2z_3^2$, then the entropy of $\sum_{k}\Pi_k\rho\Pi_k$ is
\begin{align}\nonumber
&S(\sum_k\Pi_k\rho\Pi_k)=F(\theta,\phi)\\ \nonumber
&=2-\frac{1}{4}(1+s_3\phi+\sqrt{r_3^2+2r_3c_3\phi+\theta})\log_2(1+s_3\phi+\sqrt{r_3^2+2r_3c_3\phi+\theta})\\
&-\frac{1}{4}(1+s_3\phi-\sqrt{r_3^2+2r_3c_3\phi+\theta})\log_2(1+s_3\phi-\sqrt{r_3^2+2r_3c_3\phi+\theta})\\\nonumber
&-\frac{1}{4}(1-s_3\phi+\sqrt{r_3^2-2r_3c_3\phi+\theta})\log_2(1-s_3\phi+\sqrt{r_3^2-2r_3c_3\phi+\theta})\\ \nonumber
&-\frac{1}{4}(1-s_3\phi-\sqrt{r_3^2-2r_3c_3\phi+\theta})\log_2(1-s_3\phi-\sqrt{r_3^2-2r_3c_3\phi+\theta}).
\end{align}
Note that $F(\theta, \phi)$ is an even function for the variable $\phi$, so we can focus on $\phi\in[0, 1]$
instead of $[-1, 1]$.

As in \cite{Wang2}, we compute that
\begin{align}
\begin{aligned}
&\frac{\partial F}{\partial \theta}=-\frac{1}{8}\frac{1}{\sqrt{r_3^2+2r_3c_3\phi+\theta}}\log_2\frac{1+s_3\phi+\sqrt{r_3^2+2r_3c_3\phi+\theta}}{1+s_3\phi-\sqrt{r_3^2+2r_3c_3\phi+\theta}}\\
&~~~~~~~~~~~~~-\frac{1}{8}\frac{1}{\sqrt{r_3^2-2r_3c_3\phi+\theta}}\log_2\frac{1-s_3\phi+\sqrt{r_3^2-2r_3c_3\phi+\theta}}{1-s_3\phi-\sqrt{r_3^2-2r_3c_3\phi+\theta}}<0
\end{aligned}
\end{align}

Thus $S(\sum_k\Pi_k\rho\Pi_k)=F(\theta,\phi)$ is decreasing about $\theta$. For any fixed $\phi=z_3=\phi_0\in [0, 1]$, 
the minimum of $F(\theta,\phi_0)$ should
be achieved at the maximum allowable value of $\theta$. For $\phi=\phi_0$, $\theta=c_1^2z_1^2+c_2^2z_2^2+c_3^2\phi_0^2$, and since $z_1^2+z_2^2+\phi_0^2=1$, we get that
\begin{align}\nonumber
\theta&=c_1^2z_1^2+c_2^2z_2^2+c_3^2\phi_0^2\\
&\leq c^2(z_1^2+z_2^2)+c_3^2\phi_0^2\\ \nonumber
&=c^2-c^2\phi_0^2+c_3^2\phi_0^2=c^2+(c_3^2-c^2)\phi_0^2,
\end{align}
where $c=max\{|c_1|,|c_2|\}$, and the equality can be achieved by appropriate $t, y_i$ or $z_1, z_2$.
In fact, if $|c_1|\geq |c_2|$, then $c=|c_1|$. Take $z_2=0$, then $\theta=c^2z_1^2+c_3^2\phi_0^2=
c^2(1-\phi_0^2)+c_3^2\phi_0^2=c^2+(c_3^2-c^2)\phi_0^2$. So for each fixed $\phi=\phi_0$ the maximum value of $\theta$ is $c^2+(c_3^2-c^2)\phi_0^2$.
Therefore for $\phi\in [0, 1]$ 
the minimum of $F(\theta,\phi)$, or the maximum of $2-F(\theta,\phi)$, is
given by the maximum of the following function
\begin{align}\label{e:F}
\begin{aligned}
G(\phi)&=2-F(c^2+(c_3^2-c^2)\phi^2,\phi)\\
&=\frac{1}{4}(1+s_3\phi+R_+)\log_2(1+s_3\phi+R_+)+\frac{1}{4}(1+s_3\phi-R_+)\log_2(1+s_3\phi-R_+)\\
&+\frac{1}{4}(1-s_3\phi+R_-)\log_2(1-s_3\phi+R_-)+\frac{1}{4}(1-s_3\phi-R_-)\log_2(1-s_3\phi-R_-),\\
\end{aligned}
\end{align}
where $R_{\pm}=\sqrt{r_3^2\pm 2r_3c_3\phi+c^2+(c_3^2-c^2)\phi^2}
=\sqrt{(r_3\pm c_3\phi)^2+c^2(1-\phi^2)}$.
This transforms the question of $\min S(\sum_k\Pi_k\rho\Pi_k)$ to the maximization of a one variable function.

By the definition of the one-way information deficit and Eq. (\ref{Eq2.4}) it follows that
\begin{align}\label{exact}
\begin{aligned}
&\Delta^\rightarrow(\rho)=\min_{\{B_k\}} S(\sum_k\Pi_k\rho\Pi_k)-S(\rho)\\
&=\frac{1}{4}(1-c_3+\sqrt{(r_3-s_3)^2+(c_1+c_2)^2})\log_2(1-c_3+\sqrt{(r_3-s_3)^2+(c_1+c_2)^2})\\
&+\frac{1}{4}(1-c_3-\sqrt{(r_3-s_3)^2+(c_1+c_2)^2})\log_2(1-c_3-\sqrt{(r_3-s_3)^2+(c_1+c_2)^2})\\
&+\frac{1}{4}(1+c_3+\sqrt{(r_3-s_3)^2+(c_1-c_2)^2})\log_2(1+c_3+\sqrt{(r_3-s_3)^2+(c_1-c_2)^2})\\
&+\frac{1}{4}(1+c_3-\sqrt{(r_3-s_3)^2+(c_1-c_2)^2})\log_2(1+c_3-\sqrt{(r_3-s_3)^2+(c_1-c_2)^2})\\
&-\max_{\phi\in[0,1]}G(\phi), 
\end{aligned}
\end{align}
where $c=max\{|c_1|,|c_2|\}$ in the definition
of $G(\phi)$ (see Eq. (\ref{e:F})). 
Eq. (\ref{exact}) gives an exact answer of the one-way information deficit of an arbitrary X-state
and replaces Eq. (15) of \cite{Wang}. In particular, the formulas contain cases with
nonzero components of the $z$-direction in the Bloch representation of the density matrix $\rho$
(see Remark \ref{r:comp}).

We now compute the maximum value of $G(\phi)$ $(0\leq \phi\leq 1)$ for some nontrivial cases. First we have
\begin{align}\nonumber
R_{\pm}'(\phi)&=R_{\pm}^{-1}\cdot(\pm r_3c_3+(c_3^2-c^2)\phi),\\ \label{der}
G'(\phi)&=\frac{R_+'}{4}\log_2\frac{1+s_3\phi+R_+}{1+s_3\phi-R_+}+\frac{R_-'}{4}\log_2\frac{1-s_3\phi+R_-}{1-s_3\phi-R_-},\\
&\qquad\qquad +\frac{s_3}{4}\log_2\frac{(1+s_3\phi+R_+)(1+s_3\phi-R_+)}{(1-s_3\phi+R_-)(1-s_3\phi-R_-)}.\nonumber
\end{align}

Note that
\begin{align}\label{ineq1}
&R_+^2-R_-^2=4r_3c_3\phi,\\ \label{ineq2}
&(1+s_3\phi+R_+)(1+s_3\phi-R_+)-(1-s_3\phi+R_-)(1-s_3\phi-R_-)=4(s_3-r_3c_3)\phi,\\ \label{ineq3}
&(1+s_3\phi-R_+)(1-s_3\phi+R_-)-(1+s_3\phi+R_+)(1-s_3\phi-R_-)  \\ \nonumber
&\hskip 0.5in =2(R_--R_++s_3\phi(R_++R_-))=\frac{2\phi}{R_++R_-}(-2r_3c_3+s_3(R_++R_-)^2)
\end{align}

We then have the following results:

(i) Suppose $c_3^2-c^2\geq r_3^2$, $r_3c_3\leq 0$,
and $s_3\geq 0$. Then
$R_{-}'=R_{-}^{-1}(-r_3c_3+(c_3^2-c^2)\phi)\geq 0$ for
$\phi\in [0, 1]$.
By Eq. (\ref{ineq2}), the 3rd term in $G'(\phi)$ in Eq. (\ref{der}) is nonnegative. If follows from Eq. (\ref{ineq3}) that the first $\log_2$-expression of $G'(\phi)$ is $\leq$ the second one for
$\phi\in [0, 1]$. Then we compute that
\begin{align}\nonumber
G'(\phi)&\geq \frac{R_+'}{4}\log_2\frac{1+s_3\phi+R_+}{1+s_3\phi-R_+}+\frac{R_-'}{4}\log_2\frac{1-s_3\phi+R_-}{1-s_3\phi-R_-}
\\ \label{e:1diff}
&=\frac{1}{4}\log_2\frac{1+s_3\phi+R_+}{1+s_3\phi-R_+}\left(R_+'+R_-'\log_2\frac{1-s_3\phi+R_-}{1-s_3\phi-R_-}
\left(\log_2\frac{1+s_3\phi+R_+}{1+s_3\phi-R_+}\right)^{-1}\right)\\ \nonumber
&\geq\frac{1}{4}\log_2\frac{1+s_3\phi+R_+}{1+s_3\phi-R_+}(R_+'+R_-').
\end{align}

Since $R_+'(0)+R_-'(0)=0$ and
\begin{align}\label{e:2diff}
R_+''(\phi)+R_-''(\phi)&=\frac{d}{d\phi}\left(\frac{r_3c_3+(c_3^2-c^2)\phi}{R_+}+\frac{-r_3c_3+(c_3^2-c^2)\phi}{R_-}\right)\\
&=c^2(c_3^2-c^2-r_3^2)\left(\frac1{R_+^3}+\frac1{R_-^3}\right), \nonumber
\end{align}
which is $\geq 0$, thus $G'(\phi)\geq 0$. Therefore the maximum of $G(\phi)$ on $[0, 1]$ is
\begin{align}\nonumber
G(1)&=\frac14(1+s_3+|r_3+c_3|)\log_2(1+s_3+|r_3+c_3|)\\ \nonumber
&+\frac14(1+s_3-|r_3+c_3|)\log_2(1+s_3-|r_3+c_3|)\\ \label{e:G1}
& +\frac14(1-s_3+|r_3-c_3|)\log_2(1-s_3+|r_3-c_3|)\\ \nonumber
&+\frac14(1-s_3-|r_3-c_3|)\log_2(1-s_3-|r_3-c_3|).
\end{align}
\medskip

(ii) Suppose $r_3c_3\geq 0$, $c_3^2-c^2\geq r_3^2$, and $s_3\leq 0$.
Now the 3rd term of $G'(\phi)$ in Eq. (\ref{der}) is still nonnegative, and $R_+\geq R_-$.
Note that
$R_{+}'=R_{+}^{-1}(r_3c_3+(c_3^2-c^2)\phi)\geq R_{+}^{-1}r_3c_3\geq 0$ for $0\leq \phi\leq 1$,
and the first $\log$ is $\geq$ the second $\log$ in $G'(\phi)$ by the first identity of Eq. (\ref{ineq3}). So
\begin{align}\nonumber
G'(\phi)&\geq \frac{R_+'}{4}\log_2\frac{1+s_3\phi+R_+}{1+s_3\phi-R_+}+\frac{R_-'}{4}\log_2\frac{1-s_3\phi+R_-}{1-s_3\phi-R_-}
\\ \label{e:1diff2}
&=\frac{1}{4}\log_2\frac{1-s_3\phi+R_-}{1-s_3\phi-R_-}\left(R_+'\log_2\frac{1+s_3\phi+R_+}{1+s_3\phi-R_+}
\left(\log_2\frac{1-s_3\phi+R_-}{1-s_3\phi-R_-}\right)^{-1}+R_-'\right)\\ \nonumber
&\geq\frac{1}{4}\log_2\frac{1-s_3\phi+R_-}{1-s_3\phi-R_-}(R_+'+R_-').
\end{align}
Similarly by Eq. (\ref{e:2diff}) we have $R_+''(\phi)+R_-''(\phi)\geq 0$, then
$G'(\phi)\geq 0$ for $\phi\in[0,1]$. Therefore the maximum of $G(\phi)$ on $[0, 1]$ is again $G(1)$ given in
Eq. (\ref{e:G1}).
\medskip

(iii) Suppose $s_3=r_3c_3\leq 0$, $c^2=c_3^2$, and $\max\{|c|, |r_3|\}\geq 1/\sqrt2$. 
Now Eq. (\ref{ineq2}) implies that
the third term of $G'(\phi)$ is $0$. 
Note that
$R_{+}'=R_{+}^{-1}(r_3c_3+(c_3^2-c^2)\phi)=R_{+}^{-1}r_3c_3\leq 0$.

When $c_3^2=c^2$, we have
\begin{align*}
(R_++R_-)^2&=2(r_3^2+c^2+(c_3^2-c^2)\phi^2)+2\sqrt{(r_3^2+c^2+(c_3^2-c^2)\phi^2)^2-4r_3^2c_3^2\phi^2}\\
&=2(r_3^2+c^2)+2\sqrt{(r_3^2+c^2)^2-4r_3^2c_3^2\phi^2}\\
&\geq 2(r_3^2+c^2+|r_3^2-c^2|)= 4\max\{r_3^2, c^2\}\geq 2.
\end{align*}
So $-2r_3c_3+s_3(R_++R_-)^2=s_3\left(-2+(R_++R_-)^2\right)\leq 0$, then it follows from
Eq. (\ref{ineq3}) that the first-$\log$ expression $\geq$ the second $\log$-expression in
$G'(\phi)$. Subsequently
\begin{align}\nonumber
G'(\phi)&= \frac{R_+'}{4}\log_2\frac{1+s_3\phi+R_+}{1+s_3\phi-R_+}+\frac{R_-'}{4}\log_2\frac{1-s_3\phi+R_-}{1-s_3\phi-R_-}
\\ \label{e:1diff4}
&\leq\frac{1}{4}\log_2\frac{1-s_3\phi+R_-}{1-s_3\phi-R_-}(R_+'+R_-').
\end{align}
Recall that $R_+'(0)+R_-'(0)=0$, and Eq. (\ref{e:2diff}) implies that $R_+''(\phi)+R_-''(\phi)\leq 0$, thus
$G'(\phi)\leq 0$ for $\phi\in[0,1]$. Therefore the maximum of $G(\phi)$ on $[0, 1]$ is
\begin{align}
G(0)=\frac12(1+\sqrt{r_3^2+c^2})\log_2(1+\sqrt{r_3^2+c^2})+\frac12(1-\sqrt{r_3^2+c^2})\log_2(1-\sqrt{r_3^2+c^2}).
\end{align}
\medskip

(iv) Suppose $r_3=0$. Then $R_+=R_-$. It is easy to see that
if $s_3\geq 0$ and $c_3^2\geq c^2$ (resp. $s_3\leq 0$ and $c_3^2\leq c^2$), then $G(1)$ (resp. $G(0)$)
is the maximum:
\begin{align*}
G(1)&=\frac14(1+s_3+|c_3|)\log_2(1+s_3+|c_3|)+\frac14(1+s_3-|c_3|)\log_2(1+s_3-|c_3|)\\
&\qquad+\frac14(1-s_3+|c_3|)\log_2(1-s_3+|c_3|)+\frac14(1-s_3-|c_3|)\log_2(1-s_3-|c_3|),\\
G(0)&=\frac12(1+|c|)\log_2(1+|c|)+\frac12(1-|c|)\log_2(1-|c|).
\end{align*}
\medskip

We summarize the results in Table 1.

\begin{table*}[h!]
\caption{Maximum value of $G(\phi)$.}\label{tab:table1}
    \begin{tabular}{|l|l|l|l|l||l|}\hline
        &\multicolumn{4}{|c||}{}& \\
    Cases &\multicolumn{4}{|c||}{Conditions }& Maximum\\
   \hline
    & &&&&\\
    (i) & $c_3^2-c^2\geq r_3^2$ & $r_3c_3\leq 0$ & $s_3\geq 0$ & & $G(1)$ \\
   \hline
    & &&&&\\
    (ii) & $c_3^2-c^2\geq r_3^2$ & $r_3c_3\geq 0$ & $s_3\leq 0$ & & $G(1)$ \\
   \hline
    & &&&&\\
    (iii) & $c_3^2=c^2$ &        & $s_3=r_3c_3\leq 0$ & $\max\{|r_3|, |c|\}\geq \frac{\sqrt2}{2}$ & $G(0)$ \\
   \hline
 & &&&&\\
     & $c_3^2\geq c^2$ &   & $s_3\geq 0$  &  & $G(1)$ \\
   (iv)  & &  $r_3=0$ &  &&\\
     & $c_3^2\leq c^2$ &   & $s_3\leq 0$  &  & $G(0)$ \\
    \hline
    \end{tabular}
\end{table*}

In the following we comment on our exact solutions.

\begin{remark}
In \cite{Wang2}, it was argued that if one defines $\phi=z_3$, $\theta=c_1^2z_1^2+c_2^2z_2^2+c_3^2z_3^2$,
then $\phi\in[-1,1]$, $\sqrt{\theta}\leq C=\max\{|c_1|,|c_2|,|c_3|\}$ and
the equality can be achieved.
We find that the maximum of $\theta$ is only equal to $C^2$ in degenerate cases.
In fact, the variable $\theta$ is not independent of $\phi$, and the maximum of $\theta$ should be related with $\phi$.
Actually following \cite{Wang2}, since $\theta=c_1^2z_1^2+c_2^2z_2^2+c_3^2z_3^2$, to get the maximum value $C^2$
means that $(z_1,z_2,z_3)$ is fixed,
then $\phi=z_3$ is fixed and can not vary in the whole interval $[-1,1]$.

For example, if $|c_2|>|c_1|$, $|c_2|>|c_3|$ and $\theta=C^2$, then we must have $z_1=z_3=\phi=0$ and $z_2=1$.
\end{remark}

\begin{remark}
When $r_3=s_3=0$, we have
\begin{align*}
\begin{aligned}
&G(\phi)=\frac{1}{2}(1+\sqrt{c^2+(c_3^2-c^2)\phi^2})\log_2(1+\sqrt{c^2+(c_3^2-c^2)\phi^2})\\
&~~~~~~~+\frac{1}{2}(1-\sqrt{c^2+(c_3^2-c^2)\phi^2})\log_2(1-\sqrt{c^2+(c_3^2-c^2)\phi^2}).\\
\end{aligned}
\end{align*}

Let $C=\max\{|c_1|,|c_2|,|c_3|\}$, it follows from the above table that
\begin{align*}
\max G(\phi)=\frac{1}{2}(1+C)\log_2(1+C)+\frac{1}{2}(1-C)\log_2(1-C).
\end{align*}

Therefore, when $r_3=s_3=0$,
\begin{align*}
\begin{aligned}
&\Delta^\rightarrow(\rho)=\min_{\{\Pi_k\}}S(\sum_k\Pi_k\rho\Pi_k)-S(\rho)\\
&=\frac{1}{4}(1-c_1+c_2+c_3)\log_2(1-c_1+c_2+c_3)+\frac{1}{4}(1-c_1-c_2-c_3)\log_2(1-c_1-c_2-c_3)\\
&+\frac{1}{4}(1+c_1+c_2-c_3)\log_2(1+c_1+c_2-c_3)+\frac{1}{4}(1+c_1-c_2+c_3)\log_2(1+c_1-c_2+c_3)\\
&-\frac{1}{2}(1+C)\log_2(1+C)-\frac{1}{2}(1-C)\log_2(1-C),
\end{aligned}
\end{align*}
which is the one-way information deficit of the Bell-diagonal state given in \cite{Wang}.
\end{remark}

\begin{remark}\label{r:comp} We have used $C^{++}$ programs to check our results on randomly generated sets of parameters
$c_i, r_3, s_3$ such that $\sum_i|c_i|<1$ and $|r_3|+|s_3|+|c_3|<1$. We have compared
$101\times 101$ values of $2-F(\theta, \phi)$, where
$\theta=c_1^2(1-\phi^2)\cos^2 \alpha+ c_2^2(1-\phi^2)\sin^2\alpha+c_3^2\phi^2$ at the equal
lattice points of $(\alpha, \phi)\in [0, 2\pi]\times [0, 1]$ with
$101$ values of $G(\phi)$ at the equal partition points of $\phi$ on $[0, 1]$ to find that their maximum values agree to our satisfaction.

We remark that with some simple choice of the 5 parameters such that $r_3, s_3\neq 0$, Maple fails to compute the maximum values of $G(\phi)$ and $2-F(\theta, \phi)$.

\end{remark}

\bigskip

\centerline{\bf Acknowledgments}
We thank the referee for the comments which help us improve the manuscript.
We are also indebted to Dr. Ming Liu for discussions on the topics.
The research is partially supported by
Simons Foundation grant No. 198129 and NSFC grant Nrs. 11271138 and 11531004.

\bibliographystyle{amsalpha}

\begin{thebibliography}{99}

\bibitem{HHHO} M. Horodecki, P. Horodecki, R. Horodecki, J. Oppenheim, A. Sen(De), U. Sen and B. Synak, Phys. Rev. A 71,
062307 (2005).

\bibitem{OHHH} J. Oppenheim, M. Horodecki, P. Horodecki and R. Horodecki, Phys. Rev. Lett. 89, 180402 (2002).

\bibitem{OHHH2}J. Oppenheim, K. Horodecki, M. Horodecki, P. Horodecki and R. Horodecki, Phys. Rev. A 68, 022307 (2003).

\bibitem{S} A. Streltsov, H. Kampermann and D. Bru\ss, Phys. Rev. Lett. 108, 250501 (2012).

\bibitem{C} T. K. Chuan, J. Maillard, K. Modi, T. Paterek, M. Paternostro and M. Piani, Phys. Rev. Lett. 109, 070501
(2012).

\bibitem{S2}  A. Streltsov, H. Kampermann and D. Bru\ss, Phys. Rev. Lett. 106, 160401 (2011).

\bibitem{L} S. Luo, 
Phys. Rev. A 77, 042303 (2008).

\bibitem{Wang2} Y. K. Wang, N. Jing, S. M. Fei, Z. X. Wang, J. P. Cao and F. Han,
Quantum Inf. Process. 14, 2487 (2015). 

\bibitem{Wang} Y. K. Wang, T. Ma, B. Li and Z. X. Wang, Commun. Theor. Phys. 59, 540 (2013).


\end{thebibliography}

\end{document}